\documentclass[aps,prl,twocolumn,showpacs,epsfig,longbibliography,floats]{revtex4-1}
\usepackage{amsmath}
\usepackage{color}
\usepackage{graphicx}
\usepackage{epsfig}
\usepackage{amsfonts}
\usepackage{mathrsfs}
\usepackage{mathtools}
\usepackage{amsmath}
\usepackage{float}
\usepackage{physics}
\usepackage{soul}
\usepackage{hyperref}
\hypersetup{
  colorlinks=true,
  citecolor=blue,
  linkcolor=blue,
  urlcolor=blue}

\begin{document}

\title{Fate of Quantum Scars and Chaos-Assisted Tunneling under Dissipation}

\author{Debabrata Mondal$^{1,2}$, Sohan Pati$^1$, Lea F. Santos$^3$ and S. Sinha$^1$}
\affiliation{$^1$Indian Institute of Science Education and Research-Kolkata, Mohanpur, Nadia-741246, India\\
$^2$Institut f\"ur Theoretische Physik and Center for Integrated Quantum Science and Technology, Universit\"at T\"ubingen, Auf der Morgenstelle 14, 72076 T\"ubingen, Germany\\
$^3$Department of Physics, University of Connecticut, Storrs, Connecticut 06269, USA \\	
}
\begin{abstract}
Quantum scars enable memory and recurrent dynamics, but their fate under dissipation remains poorly understood. Here, we show that distinct quantum scars can respond qualitatively differently to the same dissipative channel. We consider a collective spin model coupled to a lossy cavity and, guided by its classical limit, identify scars associated with two unstable structures: a saddle point on a dissipation-free invariant manifold with vanishing cavity field and an unstable superradiant fixed point with finite photon occupation. The former is immune to photon loss in the classical limit and exhibits persistent quantum revivals, whereas the latter couples directly to the dissipative channel, but exhibits slow dynamics. Remarkably, the superradiant scar also supports chaos-assisted quantum tunneling between two symmetry-related, macroscopically distinct semiclassical configurations, despite the presence of dissipation. Our results 
identify an experimentally accessible setting for exploring dissipative scarring and chaos-assisted tunneling in cavity QED.
\end{abstract}

\maketitle

Quantum scars are manifestations of nonergodic quantum dynamics associated with unstable classical structures. They retain memory and exhibit recurrent dynamics despite the instability of their classical counterparts.  Originally discovered as an enhanced concentration of quantum eigenfunctions around unstable classical periodic orbits~\cite{Heller1984}, quantum scarring has since been extensively investigated as a manifestation of quantum-classical correspondence in chaotic systems~\cite{Heller1984,Bogomolny1988,Berry1989,KaplanHeller1998,Kaplan1999}. More recently, scarring has acquired renewed significance with the discovery of quantum many-body scars, where atypical nonthermal eigenstates embedded in an otherwise thermalizing spectrum give rise to long-lived coherent revivals from special initial states~\cite{Rydberg_expt_scar,Abanin_Rydberg_scar2018,Papic_Rydberg_scar2018,Abanin_scars_review2021,Moessner_scars_review2022,Moudgalya_review2022, McClarty2020, Corps2026}. Establishing a connection between such anomalous many-body dynamics and classical instabilities is generally challenging. However, recent studies of systems admitting a semiclassical description have begun to bridge this gap by identifying unstable classical structures associated with quantum scarring~\cite{Abanin_PRX2020,Genuine_Scar_2,Villasenor2020,Pilatowsky2021,Mondal_TCBJJ_2022,D_Mondal_CT_Rapid,D_mondal_CT_2022,D_Mondal_KCT,Evrard2024Scar,Evrard2024UPO,Pizzi2025}.

How quantum scars are affected by coupling to an environment remains much less understood. Dissipation can profoundly modify quantum dynamics and quantum chaos~\cite{Haake_GHS, Beneti_2005, Vivek_ADD, Minganti, Deb_TCD, Prosen_PRX, Kulkarni_Dicke, Lea_Santos_GHS, Robnik_225,  Deb_OADM_2026, Mondal2026DissipationResource, Xu2021}. Nevertheless, recent studies have shown that quantum scars can persist in open systems through different mechanisms, including engineered dissipative channels that embed scar states within decoherence-free subspaces~\cite{Dissipative_scar4}, dissipative stabilization of scarred states~\cite{Jiang2026}, robustness of scarred dynamics under dephasing~\cite{Ma2025}, and non-Hermitian mechanisms that enhance many-body scarring~\cite{Shen2024}. More generally, scar-like structures have also been identified directly in Liouvillian spectra~\cite{GarciaGarcia2026,Dissipative_scar3}. Yet, how the underlying classical structure shapes the response of a quantum scar to dissipation remains largely unexplored.

Here, we show that quantum scars associated with distinct unstable classical structures can respond qualitatively differently to the same dissipative channel. We consider two interacting collective spins coupled to a lossy cavity, whose classical limit provides direct access to the structures underlying scarring. The model can also be mapped to an experimentally accessible two-component bosonic Josephson junction (BJJ) coupled to a cavity. 

In this system, we identify a scar that originates from a saddle point on a dissipation-free invariant manifold with vanishing cavity field. Because this structure is classically decoupled from photon loss, the associated quantum scar exhibits persistent revivals. A second scar originates from an unstable superradiant fixed point with finite cavity occupation and therefore couples directly to the loss channel. Its memory consequently decays, yet much more slowly than for generic chaotic initial states, with the phase-space distribution remaining anomalously localized around the unstable structure for long times. Our results show that the fate of a quantum scar under dissipation is governed by how its underlying classical structure couples to the environment.

Remarkably, the superradiant scar also supports chaos-assisted quantum tunneling~\cite{Tomsovic1994,Steck2001,Satpathi2022} despite the presence of dissipation. For small spins, mixing with surrounding chaotic states facilitates tunneling between two symmetry-related unstable branches, producing an oscillatory transfer of the quantum state between macroscopically distinct regions. The tunneling persists over a finite time window. 

{\it The model and experimental realization.--} We consider the coupled-top-Dicke model, consisting of two interacting collective spins of equal magnitude $S$ coupled to a single cavity mode of frequency $\omega_c$,
\begin{eqnarray}
    \hat{\mathcal{H}} &=& \omega_c \hat{a}^\dagger \hat{a}
    -J(\hat{S}_{1x}+\hat{S}_{2x})
    +\frac{V}{S}\hat{S}_{1z}\hat{S}_{2z} \nonumber\\
    &&+\frac{\lambda}{\sqrt{2S}}
    (\hat{S}_{1z}+\hat{S}_{2z})(\hat{a}+\hat{a}^\dagger),
    \label{CTD_Model}
\end{eqnarray}
where $\lambda$ is the spin-cavity coupling. The second and third terms
describe the coupled-top model, while the last term couples the two collective spins to the cavity field, yielding a two-spin
generalization of the Dicke model
\cite{Dicke1954,Zhiqiang2017,ScottPerkins2014,Esslinger2021,
EmaryBrandes2003,Altland2012,Ciuti2014,Chitra2018,ScottPerkins2020}.

A possible physical realization of the coupled-top-Dicke model is provided by a two-component BJJ coupled to an optical cavity, as illustrated in Fig.~\ref{Fig1}(a) \cite{Wang2010}. This setting combines experimentally realizable BJJs with cavity-mediated coupling of atomic degrees of freedom~\cite{Oberthaler_1,Landig2016}. For each bosonic species, the Schwinger mapping
\begin{equation}
    \hat S_{ix}
    =\frac{1}{2}
    (\hat b^\dagger_{iL}\hat b_{iR}
    +\hat b^\dagger_{iR}\hat b_{iL}),
    \qquad
    \hat S_{iz}
    =\frac{1}{2}(\hat n_{iL}-\hat n_{iR})
\end{equation}
defines a collective spin of magnitude $S=N/2$, where $N$ is the number of bosons of each species. The corresponding BJJ-cavity Hamiltonian is
\begin{eqnarray}
\hat{\mathcal H}_{\rm BJJ-cav}
&=& \omega_c\hat a^\dagger\hat a
+\sum_{i,\ell}\left[
-\frac{J}{2}\hat b^\dagger_{i,\ell}\hat b_{i,\bar\ell}
+\frac{V}{2N}\hat n_{i,\ell}\hat n_{\bar i,\ell}
\right]
\nonumber\\
&&+\sum_i\frac{\lambda}{2\sqrt N}
(\hat a+\hat a^\dagger)
(\hat n_{iL}-\hat n_{iR}),
\label{BJJ_cavity}
\end{eqnarray}
where $i=1,2$ labels the two species and $\ell=L,R$ the two wells, with $\bar\ell$ and $\bar i$ denoting the opposite species and well, respectively~\cite{Gati2006,Oberthaler_1,Oberthaler_2, Mondal_TCBJJ_2022,Landig2016}.
We set $\hbar=1$ and measure energy (time) in units of $J$ ($1/J$).

In the presence of photon loss with rate $\kappa$, the non-unitary evolution of the density matrix is described by the Lindblad master equation \cite{Gorini1976, Lindblad1976, Breuer2007},
\begin{equation}
\dot{\hat{\rho}}=-i[\hat{\mathcal{H}}, \hat{\rho}]+\frac{\kappa}{2}\left(2\hat{a} \hat{\rho} \hat{a}^\dag-\left\{\hat{a}^\dag \hat{a}, \hat{\rho}\right\}\right),
\label{Lindblad_Master_Equation}
\end{equation}
which indicates that the environment couples directly only to the
cavity mode [Fig.~\ref{Fig1}(a)]. 

\begin{figure}[t]
\centering
\includegraphics[scale=0.15]{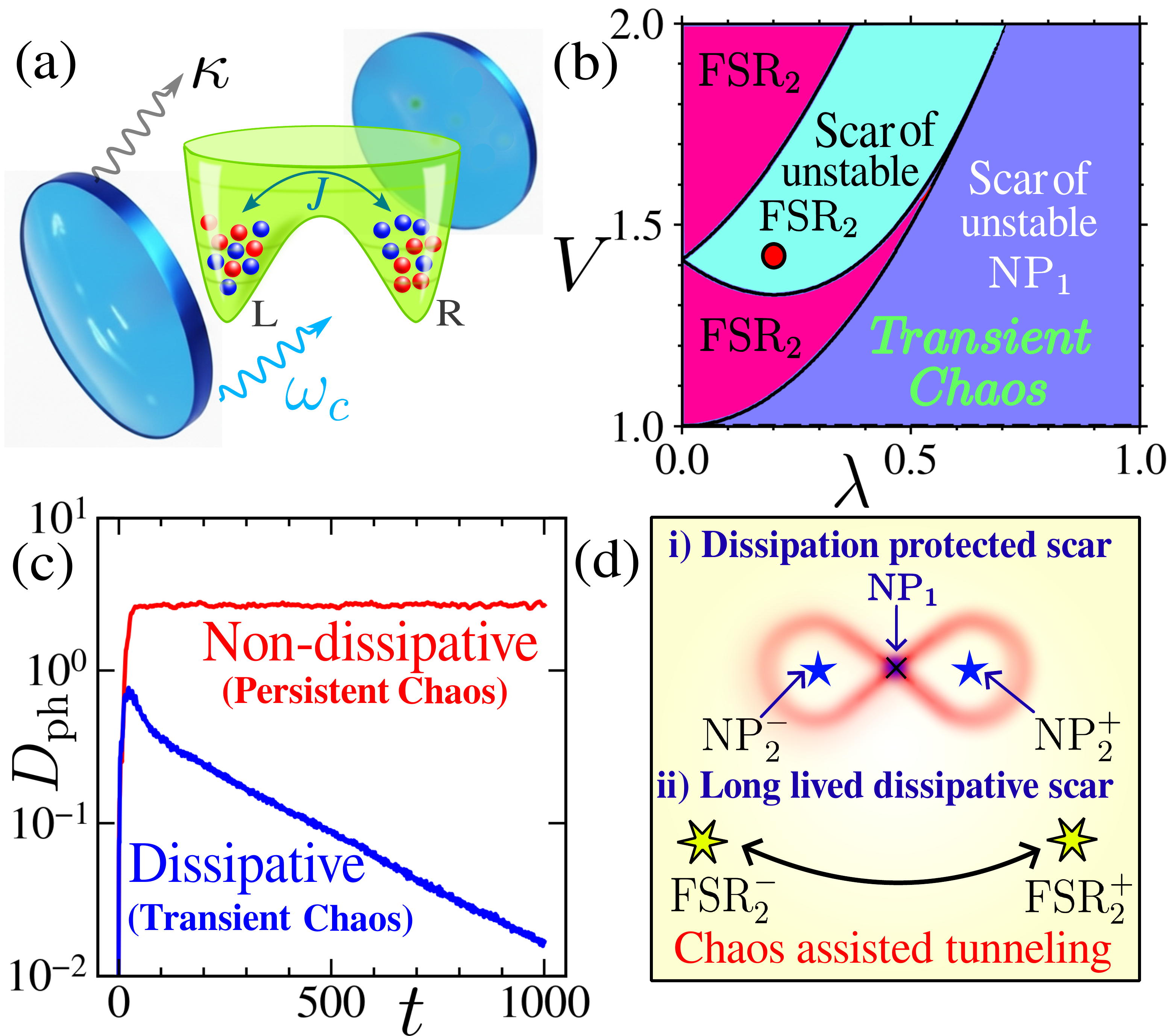}
\caption{(a) Schematic of a two-component Bose-Josephson junction (BJJ) coupled to a cavity of frequency $\omega_c$ and photon loss rate $\kappa$. (b) Phase diagram in the $\lambda-V$ plane for $\kappa=0.1$, showing the regions associated with the normal fixed point NP$_1$ and the superradiant fixed point FSR$_2$. The FSR$_2$ point is unstable in the cyan region, and the red dot marks the parameters used to study its quantum scar in Fig.~\ref{Fig3}. Transient chaos occurs in the entire region. (c) Dynamics of the photonic decorrelator $D_{\rm ph}$ exhibiting transient chaos at $\lambda=0.5,V=1.5$. (d) Schematic of the classical structures underlying the two quantum scars: NP$_1$ is an unstable saddle point in the regular regime with separatrix enclosing the stable points NP$_2^\pm$, whereas FSR$_2$  is an unstable fixed point embedded in a transiently chaotic region. The former supports dissipation-protected scarring, while the latter is associated with dissipative scarring and chaos-assisted tunneling.
}
\label{Fig1}
\end{figure}

In the limit $S\rightarrow \infty$, the scaled cavity field 
 $\alpha = \langle \hat{a}\rangle/\sqrt{S} = (x+i p)/\sqrt{2}=\sqrt{n}\exp(i\psi)$, with scaled photon number $n=|\alpha|^2$ and phase $\psi$, and the normalized spin vectors $\vec{s}_i = \langle\hat{\vec{S}}_i\rangle/S$ can be treated as classical variables~\cite{Souza2023, Exactness_1, Exactness_2}. The corresponding classical equations of motion, obtained from the Lindblad master equation in the mean-field limit, are
\begin{subequations}
	\begin{eqnarray}
		\dot{\alpha} &=& -\left(\frac{\kappa}{2}+i\omega_c\right)\alpha
		- i\frac{\lambda}{\sqrt{2}}(s_{1z}+s_{2z}) ,
        \label{Eq:alphaDot}
        \\
		\dot{s}_{i+} &=& i \frac{\lambda}{\sqrt{2}}(\alpha+\alpha^*) s_{i+}
		+ i V s_{i+} s_{\bar{i}z}
		+ i J s_{iz} , \\
		\dot{s}_{iz} &=& \frac{iJ}{2}(s_{i+} - s_{i-}).
	\end{eqnarray}
\end{subequations}
Each spin lies on the unit Bloch sphere, $\vec{s}_i=(\sin\theta_i\cos\phi_i,\sin\theta_i\sin\phi_i,\cos\theta_i)$, with canonical variables $\phi_i$ and $z_i=\cos\theta_i$. In the BJJ realization, $z_i=(n_{iL}-n_{iR})/N$ is the population imbalance and $\phi_i$ is the relative phase between the two wells. These classical equations allow us to identify the unstable structures underlying the quantum scars and, crucially, to determine their coupling to the dissipative cavity mode. 

For the quantum dynamics, we evolve initial product coherent states
using the stochastic wave-function formalism and reconstruct the density matrix from an ensemble of quantum
trajectories \cite{Daley_2014,Carmichael1993,Molmer1993}.

{\it Invariant manifolds and dissipation-free dynamics.—} The classical equations of motion admit two invariant manifolds associated with the quantum scars investigated below. Introducing the symmetric and antisymmetric spin combinations $s_{\pm a}=(s_{1a}\pm s_{2a})/2$, with $a=x,y,z$, these manifolds are defined by
\begin{eqnarray}
&\!\!\!\!\!\text{Symmetric:} &\, \{x,p\ne0,s_{x-}=s_{y-}=s_{z-}=0\}.\quad\qquad\label{Symmetric_class} \\
	&\!\!\!\!\!\text{Anti-symmetric:} & \,\{x=p=s_{x-}=s_{y+}=s_{z+}=0\},\quad \label{Anti_symmetric_class}
\end{eqnarray}
%
%

The symmetric manifold is defined by identical spin configurations, $\vec{s}_1=\vec{s}_2$. Its invariance follows from the exchange symmetry $1\leftrightarrow2$ of the equations of motion. Two spins initialized in the same configuration remain identical throughout the evolution. 
The cavity field is unrestricted and can therefore couple to photon loss. 

The antisymmetric manifold is characterized by $s_{1x}=s_{2x}$, $s_{1y}=-s_{2y}$, $s_{1z}=-s_{2z}$, and $\alpha=0$. These conditions are preserved by the classical equations of motion. 
In particular, $s_{1z}+s_{2z}=0$ eliminates the source term for the cavity field in Eq.~(\ref{Eq:alphaDot}), which consequently remains zero and is not generated dynamically. 
Photon loss therefore has no effect on trajectories confined to this manifold, leading to a dissipation-free dynamics equivalent to the dynamics of an effective Lipkin-Meshkov-Glick model~\cite{LMG}. 
%

{\em Classical structures underlying the quantum scars.--} We now identify the classical fixed points $\mathbf{Q}^* = \{x^*,p^*,s_{ix}^*,s_{iy}^*,s_{iz}^*\}$ relevant to the two quantum scars  investigated below. 

The first is the normal fixed point NP$_1$, where $s_{ix}^*=1$ and $s_{iy}^*=s_{iz}^*=0$.  Using the canonical spin variables, this fixed point is located at
\begin{equation}
{\rm NP}_1: \quad z_i^*=0, \,\, \phi_i^*=0.
\end{equation}
NP$_1$ belongs to the antisymmetric dissipation-free subspace introduced above, so $x^*=p^*=0$. For $V<J$, the point is stable. At $V=J$, it undergoes a bifurcation and, for $V>J$, it persists as an unstable saddle point while the two following symmetry-related antiferromagnetic fixed points emerge with $s_{ix}^*=J/V$, $s_{iy}^*=0$ and $s_{1z}^* =-s_{2z}^*=\pm \sqrt{1-\left(\frac{J}{V}\right)^2}$. In canonical variables, the two branches are
\begin{equation}
{\rm NP}_2^\pm: \quad z_1^*=-z_2^*
=\pm\sqrt{1-\left(\frac{J}{V}\right)^2}, \,\, \phi_i^*=0.
\end{equation}

In the parameter regime considered below, with $V>J$ and moderate spin-cavity coupling $\lambda$, the unstable NP$_1$ saddle point lies on a homoclinic separatrix whose two lobes enclose the stable NP$_2^\pm$ fixed points in the $(z_i,\phi_i)$ phase space [Fig.~\ref{Fig1}(d)]. This unstable classical
structure underlies the first quantum scar. 

The second classical structure relevant to scarring is the superradiant fixed point FSR$_2$, which belongs to the symmetric manifold, $\vec{s}_1=\vec{s}_2$, and has a finite cavity field. Its two
symmetry-related branches are given by
\begin{eqnarray}
{\rm FSR}_2^{\pm}: \,\,z_{1}^*\!\!&=&\!\!z_{2}^*=\pm\sqrt{1-\frac{J^{2}(\kappa^2 +  4\omega_c^{2})^2}{\left[8 \omega_c\lambda^{2}-(\kappa^2 +  4\omega_c^{2})V\right]^{2}}},\qquad\nonumber\\
x^*\!\!&=&\!\!-\frac{8\omega_c \lambda}{(\kappa^2 +  4\omega_c^{2})} z_{1}^*, \,\, p^* = \frac{\kappa x^*}{2\omega_c},\,\, \phi_{i}^*=\pi, \quad
\label{FSR1}
\end{eqnarray}	
The stability of FSR$_2$ depends on the model parameters. In the isolated system, stable FSR$_2$ fixed points are surrounded by regular islands embedded in a chaotic sea. As the parameters are varied, these islands disappear and FSR$_2$ becomes unstable (see Supplemental Material (SM)~\cite{footSM}). We focus on this unstable regime, in which the two symmetry-related branches of FSR$_2$ are associated with chaotic dynamics. 
Even in the presence of dissipation, FSR$_2$ is unstable in this regime, which underlies the second dissipative quantum scar and the chaos-assisted tunneling between the two branches discussed below [Fig.~\ref{Fig1}(d)]
%

Unlike NP$_1$, FSR$_2$ has a nonzero cavity field and is therefore directly affected by photon loss. For $\kappa>0$, dissipation suppresses chaotic dynamics at long times, giving rise to transient chaos (SM~\cite{footSM}). This behavior is illustrated by the photonic decorrelator~\cite{Chatterjee2020DuffingOTOC,Ruidas2024SemiclassicalMIPT},
\begin{equation}
D_{\rm ph}(t)=
\overline{\sqrt{[x^a(t)-x^b(t)]^2+[p^a(t)-p^b(t)]^2}},
\end{equation}
where $a$ and $b$ denote nearby initial conditions and the overline indicates an average over initial configurations. Its initial exponential growth followed by decay [Fig.~\ref{Fig1}(c)] is characteristic of the crossover from transient chaotic spreading to
regular long-time dynamics.


\begin{figure*}[t]
 \centering
\includegraphics[width=\linewidth]{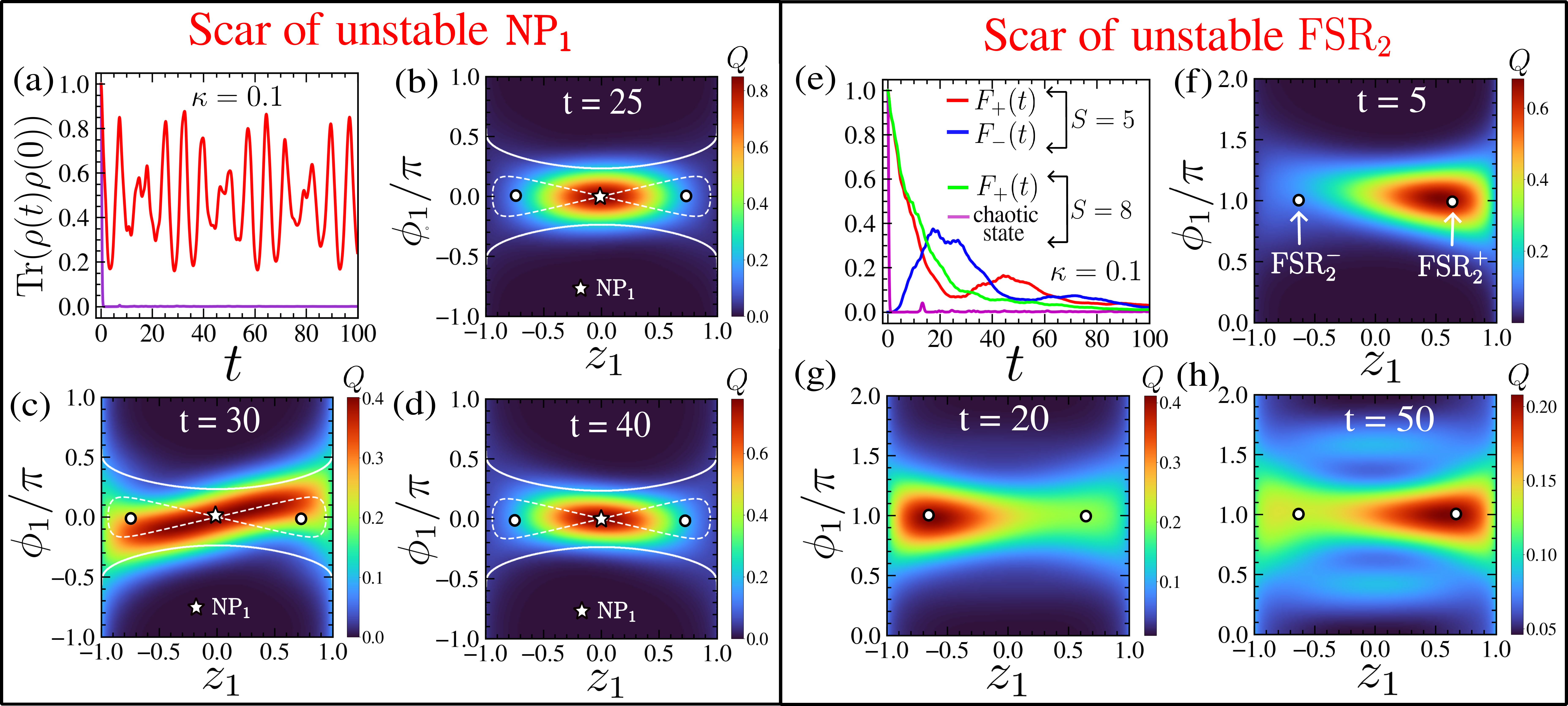}
\caption{ \textbf{ Quantum scarring phenomena of unstable NP$_1$}: (a) Dynamics of survival probability ${\rm Tr}(\rho(t)\rho(0))$  and (b,c,d) Husimi distribution $Q$ of reduced density matrix of one of the spins for three different times at $V=1.5,\lambda=0.5$. The purple and red lines correspond to the initial state representing a generic phase space point and unstable NP$_1$, respectively.\\
\textbf{Quantum scarring phenomena of unstable FSR$_2$}: (e) Dynamics of the overlaps $F_{j}(t)$ corresponding to two branches of unstable FSR$_2^{j}$  [see the main text], where (red, green) lines for $j=+$ and blue line for $j=-$. The initial state is localized at one of the branch FSR$_2^+$. $F_+(t)$ corresponds to the survival probability and $F_-(t)$ is the overlap with the opposite branch FSR$_2^-$.
%
%
The green color line corresponds to $S = 8$. The purple line represents the survival probability starting from a generic phase space point. (f,g,h) Husimi distribution $Q$ of reduced density matrix of one of the spins at three different times. Coupling strengths for panels (e-h) is $V=1.4,\lambda=0.2$ (see Fig.\ref{Fig1}(b)). Parameter chosen: $\kappa=0.1, S=5$ for all panels. in (e), if I understood the order in the text correctly, I think the legend should be in this order: $F_{+}(t)$ (green) and chaotic (purple) for $S=8$; then $F_{+}(t)$ (red) and $F_{-}(t)$ (blue) for $S=5$ 
}
	\label{Fig3}
\end{figure*}

We now investigate the quantum dynamics associated with the two unstable classical structures identified above, NP$_1$ and FSR$_2$, and their distinct responses to photon loss.

{\it (i) Dissipation-protected scar of NP$_1$.--} We first initialize the system in a spin-cavity coherent state centered at the unstable saddle point NP$_1$, $\hat{\rho}(0)=\ket{\Psi_{\rm NP_1}}\bra{\Psi_{\rm NP_1}}$. Because photon loss generally drives an initially pure state into a mixed state, we study the evolution of the survival probability
\begin{equation}
    F(t)={\rm Tr}\!\left[\hat{\rho}(t)\hat{\rho}(0)\right],
\end{equation}
where $\hat{\rho}(t)$ is reconstructed from an ensemble of quantum trajectories. The survival probability exhibits recurrent oscillations with negligible decay [red line in Fig.~\ref{Fig3}(a)], demonstrating persistent memory of the initial state despite photon loss. This behavior contrasts with the rapid decay observed for an initial coherent state centered at a generic point in the chaotic region [purple line in Fig.~\ref{Fig3}(a)]. 

The phase-space dynamics reveals the origin of these revivals. The Husimi distribution, initially localized at NP$_1$, spreads along the homoclinic separatrix and subsequently reconcentrates near the unstable saddle point [Figs.~\ref{Fig3}(b)--\ref{Fig3}(d)]. These recurrent returns to NP$_1$ account for the oscillations of $F(t)$ and provide the phase-space signature of quantum scarring.

The persistence of these revivals under photon loss follows from the dissipation-free character of the underlying classical structure.
As established above, the homoclinic separatrix is confined to the antisymmetric manifold of two spins, where the cavity remains empty and photon loss does not affect the classical dynamics. The resulting long-lived quantum recurrences therefore constitute a dissipation-protected quantum scar~\cite{Dissipative_scar4}.

{\it (ii) Dissipative scar of FSR$_2$ and chaos-assisted tunneling.--} We now initialize the system in a spin-cavity coherent state $\ket{\Psi_{\rm FSR_2^{+}}}$ centered at one of the two symmetry-related branches of the unstable fixed point FSR$_2$. In the presence of photon loss, the survival probability $F_{+}(t) = {\rm Tr}\left(\hat{\rho}(t) \ket{\Psi_{\rm FSR_2^{+}}} \bra{\Psi_{\rm FSR_2^{+}}}\right)$ decays slowly [green line in Fig.~\ref{Fig3}(e)], rather than exhibiting the revivals observed in the isolated system (SM~\cite{footSM}). 
Yet, the decay is significantly slower than for a generic initial state in the chaotic region [purple line in Fig.~\ref{Fig3}(e)], revealing a dissipative scar. Consistently, for sufficiently large $S$, the Husimi distribution remains localized near FSR$_2$ over an extended time window before gradually spreading into the surrounding phase space~\cite{footSM}. Unlike the NP$_1$ scar, however, the FSR$_2$ scar is not protected from photon loss.

For small $S$, the dynamics reveals an additional phenomenon. Instead of a monotonic slow decay, $F_{+}(t)$ develops oscillations superimposed on an overall dissipative decay [red line in Fig.~\ref{Fig3}(e)]. The overlap with the coherent state centered at the opposite branch, $F_{-}(t)= {\rm Tr}\left(\hat{\rho}(t) \ket{\Psi_{\rm FSR_2^{-}}} \bra{\Psi_{\rm FSR_2^{-}}}\right)$, exhibits complementary oscillations [blue line in Fig.~\ref{Fig3}(e)], indicating quantum tunneling between the two macroscopically distinct FSR$_2$ configurations. The Husimi distributions in Figs.~\ref{Fig3}(f)--\ref{Fig3}(h) confirm this behavior, showing the phase-space distribution alternating between the two symmetry-related unstable FSR$_2$ branches.

This macroscopic quantum tunneling is connected to the instability of FSR$_2$ and the surrounding chaotic dynamics. As shown in SM~\cite{footSM} for the isolated system, when FSR$_2$ is stable, tunneling between its two branches  is strongly suppressed, while mixing with the surrounding chaotic
states facilitates transfer between the two branches, identifying the process as chaos-assisted macroscopic quantum tunneling~\cite{Tomsovic1994,Steck2001,Arnal2020,Satpathi2022}. Remarkably, this tunneling survives in the presence of photon loss over a finite time window. At longer times, dissipation-induced spreading of the phase-space distribution suppresses the coherent transfer and leads to the decay of the survival probability.

{\it Discussion.--}
Our results show that quantum scarring can persist under dissipation, but its robustness depends on how the underlying classical structures couple to the environment. A scar associated with a dissipation-free invariant manifold exhibits persistent quantum revivals, whereas a scar associated with an unstable superradiant fixed point retains memory of its initial state over a finite time window. We further demonstrate that chaos-assisted macroscopic quantum tunneling between two symmetry-related configurations can survive photon loss, although dissipation eventually suppresses the coherent transfer. These findings highlight the role of classical phase-space structures in determining the persistence of nonergodic quantum dynamics in open systems.

{\it Acknowledgments.--} D.M. acknowledges support from the Alexander von Humboldt Foundation through a Humboldt Research Fellowship. D.M and S.S thank G. Vivek for useful discussions. L.F.S. acknowledges support from the U.S. Department of Energy under Grant No. DE-SC0026657.

\bibliography{bibliography_V5.bib}


\onecolumngrid

\vspace*{0.4cm}

\begin{center}
{\large \bf Supplemental Material: 
\\Fate of Quantum Scars and Chaos-Assisted Tunneling under Dissipation}\\
\vspace{0.6cm}
Debabrata Mondal$^{1,2}$, Sohan Pati$^1$, Lea F. Santos$^3$ and S. Sinha$^1$\\
$^1${\it Indian Institute of Science Education and Research-Kolkata, Mohanpur, Nadia-741246, India}\\
$^2${\it Institut f\"ur Theoretische Physik and Center for Integrated Quantum Science and Technology, Universit\"at T\"ubingen, Auf der Morgenstelle 14, 72076 T\"ubingen, Germany}
\\
$^3${\it Department of Physics, University of Connecticut, Storrs, Connecticut 06269, USA }	
\end{center}

\maketitle


\renewcommand{\theequation}{S\arabic{equation}}
\renewcommand{\thefigure}{S\arabic{figure}}

This supplemental material provides additional figures and analyses that support the discussions in the main text.

\section{Derivation of open coupled-top-Dicke model and classical dynamics}
As discussed in the main text, the ``open coupled-top Dicke” (OTCD) model, which describes two interacting large spins coupled to a cavity mode, can be realized using a cavity-coupled two-component Bosonic Josephson junction (BJJ). The corresponding system is described by a Bose–Hubbard dimer coupled to a cavity with the Hamiltonian \cite{Landig2016}
\begin{eqnarray}
	\hat{\mathcal{H}}=\omega_c \hat{a} ^\dagger \hat{a} -\frac{J}{2} \sum_{i=1,2}\left( \hat a^\dagger_{iL}\hat a_{iR} + \hat a^\dagger_{iR}\hat a_{iL} \right)+
	\frac{V}{N} \left( \hat n_{1L}\hat n_{2L} + \hat n_{1R}\hat n_{2R} \right) + \frac{\lambda}{2\sqrt{N}}\sum_{i=1,2}(\hat{a} + \hat{a}^\dagger)(\hat{n}_{i{\rm L}} - \hat{n}_{i{\rm R}}),
\end{eqnarray}
where the number of bosons for each species with index $i=1,2$ are equal and is given by $N = \hat{n}_{iL}+\hat{n}_{iR}$.
Within the Schwinger boson representation,
$
\hat{S}_{i-}= \hat{a}_{i{\rm R}}^{\dagger}\hat{a}_{i{\rm L}}, \,\, \hat{S}_{iz}=(\hat{n}_{i{\rm L}}-\hat{n}_{i{\rm R}})/2,
$
this BJJ can be mapped to the system of two interacting large spins of equal magnitude $S=N/2$ (see Eq.1 of the main text), given by,
\begin{eqnarray}
	\hat{\mathcal{H}} = \omega_c \hat{a} ^\dagger \hat{a}  -J(\hat{S}_{1x} + \hat{S}_{2x})+ \frac{V}{S} \hat{S}_{1z} \hat{S}_{2z} + \frac{\lambda}{\sqrt{2S}}(\hat{S}_{1z} + \hat{S}_{2z})(\hat{a} + \hat{a} ^\dagger),
	\label{CTD_Model}	
\end{eqnarray}
apart from the zero point energy term.

In this section, we discuss the classical dynamics and the various non-equilibrium phases of the open coupled-top-Dicke (OCTD) model in the presence of photon loss.
The non-unitary evolution of the system's density matrix $\hat{\rho}$ is governed by the Lindblad master equation, as given in Eq.~(4) of the main text. The time evolution of the expectation value of any operator $\hat{A}$ is obtained using the equation 
$\frac{d\langle \hat{A} \rangle }{ dt} = 
{\rm Tr}(\hat{A} \dot{\hat{\rho}})$.
In the limit $S\rightarrow \infty$, the scaled operators $\hat{a}/\sqrt{S} = (\hat{x} + \iota \hat{p})/\sqrt{2}$ and $\hat{\vec{s}}_i = \hat{\vec{S}}_i/S$ behave classically, since they satisfy the commutation relations $[\hat{x}_i,\hat{p}_i]= \iota/S$ and $[\hat{s}_{ia}, \hat{s}_{jb }] =\delta_{ij} \iota \epsilon_{abc} \hat{s}_{ic }/S$, where $1/S$ plays the role of a reduced Planck constant.
The classical equations of motion (EOM) for these collective variables can be derived from the master equation using the decoupling approximation $\langle \hat{A}\hat{B}\rangle\approx\langle \hat{A}\rangle \langle\hat{B}\rangle$ and are given by,
\begin{subequations}
	\begin{eqnarray}
		\dot{\alpha} &=&  -\left(\frac{\kappa}{2}+i\omega_c\right)\alpha -i\frac{\lambda}{\sqrt{2}}(s_{1z}+s_{2z})\\
		\dot{s}_{ix} &=& - \frac{\lambda}{\sqrt{2}} (\alpha+\alpha^*) s_{iy} - V s_{iy}s_{\bar{i}z}, \\
		\dot{s}_{iy} &=& \frac{\lambda}{\sqrt{2}} (\alpha+\alpha^*) s_{ix} + V s_{ix}s_{\bar{i}z} + J s_{iz} , \\
		\dot{s}_{iz} &=& -J s_{iy},
	\end{eqnarray}
	\label{EOM}
\end{subequations} 
where the classical photon field is represented by $\alpha = \langle \hat{a}\rangle/\sqrt{S} = (x+i p)/\sqrt{2}=\sqrt{n}\exp(i\psi)$ with number $n=|\alpha|^2$ and phase $\psi$. The scaled spin vector can be written as $\vec{s}_i = (\sin\theta_i\cos\phi_i,\sin\theta_i\sin\phi_i,\cos\theta_i)$, where $s_{iz}=\cos\theta_i$ and $\phi_i$ are conjugate variables.

\textbf{Fixed points and their stability:} The steady states of the coupled-top-Dicke model are characterized by the fixed points $\mathbf{Q}^*=\{x^*,p^*,s_{ix}^*,s_{iy}^*,s_{iz}^*\}$ of the EOM [Eq.\eqref{EOM}], obtained by setting $\dot{\mathbf{Q}}=0$.  
The stability of each fixed point is determined by linearizing the EOM around $\mathbf{Q}^*$. A small fluctuation $\delta \mathbf{Q}$ from these fixed points evolves as $\delta \mathbf{Q}(t)=\delta \mathbf{Q} e^{\lambda t}$, where $\lambda$ are the complex eigenvalues of the stability matrix~\cite{Strogatz}. 

\begin{figure}[t]
	\centering
	\includegraphics[width=0.6\linewidth]{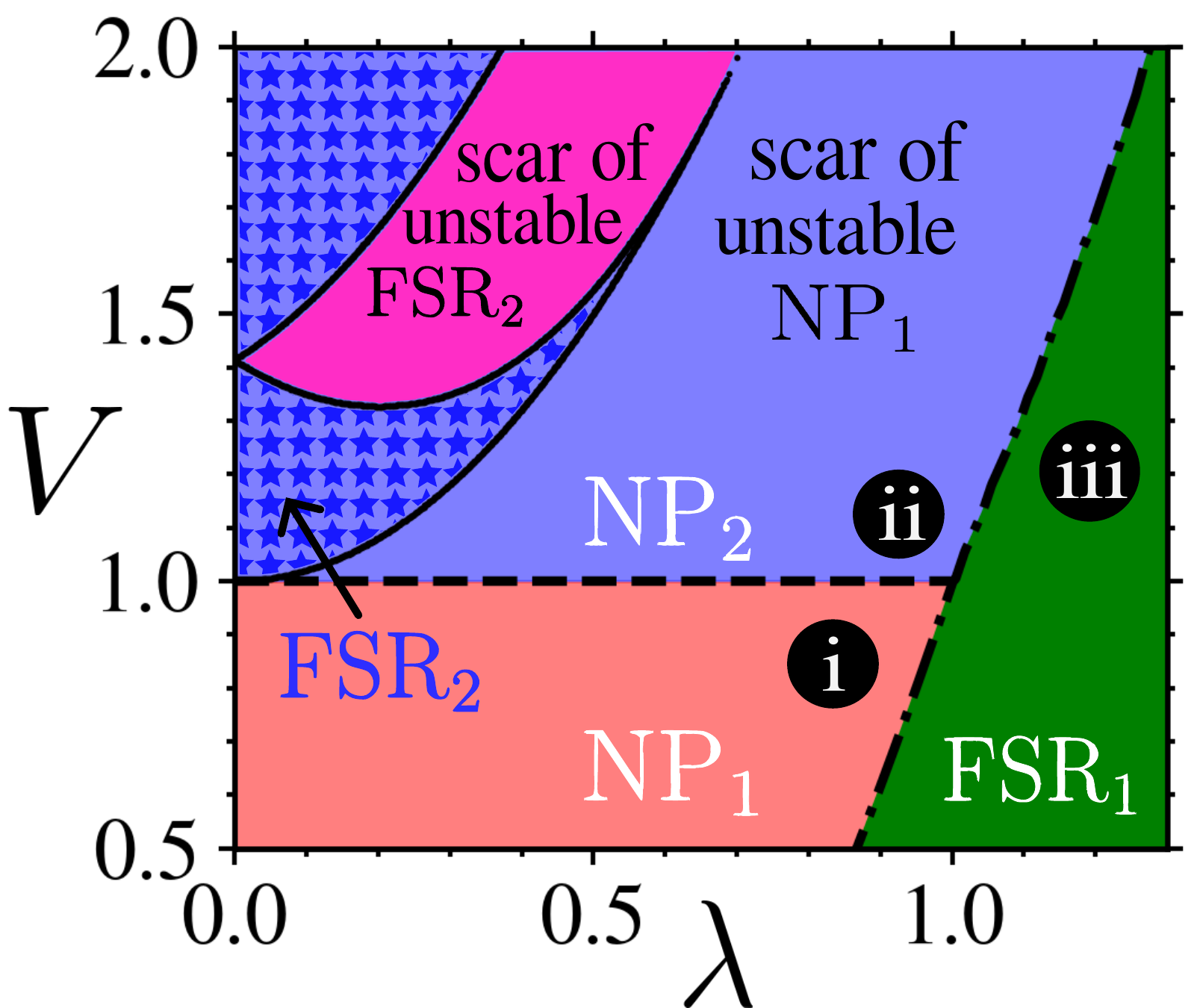}
	\caption{ The classical phase-diagram in $V-\lambda$ plane for $\kappa=0.1$, summarizing different dynamical regimes and fixed points. The stable fixed points in regions (i), (ii), (iii) are NP$_1$, NP$_2$ and FSR$_1$ respectively. The stable regime of FSR$_2$ for the isolated case $\kappa=0$ is marked by blue star-filled area. The scar of unstable FSR$_2$ occurs in pink region. The scar of unstable NP$_1$ occurs in region ii.
	}
	\label{FigS1A}
\end{figure}

A fixed point is a stable attractor if $\mathrm{Re}(\lambda) < 0$ for all eigenvalues, in which case nearby trajectories converge to it. In contrast, for a Hamiltonian system (isolated case), the fixed points are center as $\mathrm{Re}(\lambda) = 0$ for all eigenvalues, supporting periodic orbits around them.
Unlike these two scenarios, in the present case, for all the fixed points, the real parts of only two eigenvalues are negative, while the remaining ones are zero. Consequently, the fluctuations of certain normal modes vanish, whereas the others persist as oscillations with frequency $\tilde{\omega} = \mathrm{Im}(\lambda)$.
This feature makes the fixed points partial attractors, where the dynamics is attracted towards them, but instead of converging to these fixed points, it eventually forms an orbit in a certain reduced phase space. Note that these oscillatory dynamics are different from limit cycles, as the orbits are not unique and the amplitude of oscillation depends on the initial fluctuation.
Different fixed points such as normal phases NP$_1$, NP$_2$ and superradiant phases FSR$_1$, FSR$_2$ [see main text for details] and their stability regimes  are summarized in Fig.\ref{FigS1A}.

\begin{figure}[t]
	\centering
	\includegraphics[width=\linewidth]{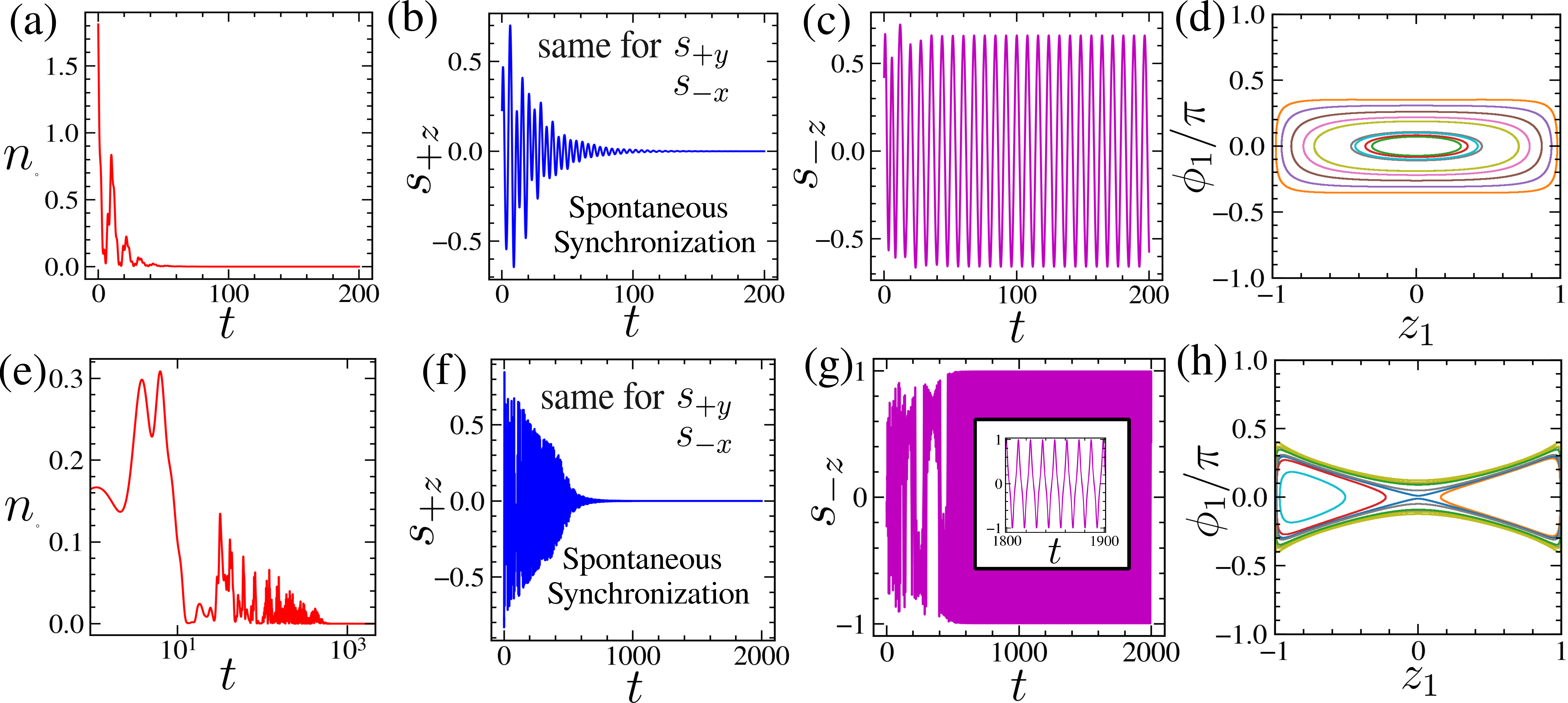}
	\caption{ Classical dynamics in region-i,ii of the phase diagram in Fig.1(b) of the main text. The dynamics of (a,e) photon number $n$, collective spin variables (b,f) $s_{+z}$, (c,g) $s_{-z}$ and (d,h) the phase-portrait in $z_1-\phi_1$ corresponding to one of the spins for region-i,ii at $(V=0.5,\lambda=0.5), (V=1.8,\lambda=0.2)$, respectively. The timescale of synchronization is large for region-ii compared to region-i due to the transient chaos. The inset of panel (g) clearly shows oscillatory behavior at long time. The phase portrait in the (d,h) plane is plotted over a long time interval after discarding the transient dynamics. Other parameters chosen for all the panels: $\omega_c=1,J=1,\kappa=0.3$.
	}
	\label{FigS1}
\end{figure}

{\bf Anti-symmetric dynamical class and decoherence-free subspace:}
Next, we discuss the emergence of periodic orbits and spontaneous synchronization between the two spins [see the details in the main text]. As mentioned in the main text, the classical EOM (Eq.\eqref{EOM}) of the OCTD model are invariant under the exchange of two spins $(S_1 \leftrightarrow S_2)$, which plays a crucial role in the formation of coherent oscillatory phase.
To understand the effect of this exchange symmetry on the dynamics, we rewrite the EOM in terms of the new classical variables $s_{\pm a} = (s_{1a} \pm s_{2a})/2\,\,(a=x,y,z)$, which are given by,
\begin{subequations}
	\begin{align}
		\dot{x} &= \omega p - \tfrac{\kappa x}{2}, \\
		\dot{p} &= -2\lambda s_{+z} - \omega x - \tfrac{\kappa p}{2}, \\
		\dot{s}_{\pm x} &= - \lambda x s_{\pm y} - V s_{\pm y}s_{+z} + V s_{\mp y}s_{-z}, \\
		\dot{s}_{\pm y} &= \lambda x s_{\pm x} + V s_{\pm x}s_{+z} -V s_{\mp x}s_{-z} + J s_{\pm z} , \\
		\dot{s}_{\pm z} &= -J s_{\pm y}.
	\end{align}
	\label{EOM_dynamical_class}
\end{subequations}
It is evident from the above EOM (Eq.\eqref{EOM_dynamical_class}) that either of the following conditions 
\begin{eqnarray}
	&&\!\!\!\!\!\text{(I) anti-symmetric} : \{x=p=s_{x-}=s_{y+}=s_{z+}=0\}\quad \label{Anti_symmetric_class}\\
	&&\!\!\!\!\!\text{(II) symmetric}: \{x,p\ne0,s_{x-}=s_{y-}=s_{z-}=0\},\quad\qquad\label{Symmetric_class}
\end{eqnarray}
can be satisfied, leading to the constraint dynamics in a reduced phase space. 
The fixed points and the corresponding dynamics can be classified into the above-mentioned categories, each satisfying its respective constraints. However, in general, the dynamics need not remain confined to these classes.
The anti-symmetric dynamical class corresponds to the vanishing photon field, which determines the asymptotic dynamics of spins in the normal phase (parameter region I,II of the phase diagram in Fig.(1b) of the main text) as the photon number vanishes at long time due to dissipation.
%
Importantly, the constraint dynamics on the anti-symmetric dynamical class leads to the synchronization between the two spins $s_{x-}=0,s_{y+}=0,s_{z+}=0$, irrespective of initial conditions, as depicted in Fig.\ref{FigS1}(b,f). 
In the context of the Bosonic Josephson junction, the relative phase $\phi_i$ and the inter-well population imbalance $z_i$ of the two species of bosons satisfy the constraints $\phi_+=(\phi_1+\phi_2)/2=0$, $z_+=(z_1+z_2)/2=0$, respectively.
Moreover, the dynamics within the class-I reduces to an effective dissipation-free motion of spins since the photon field is zero, leading to a coherent oscillatory motion, as illustrated in Fig.\ref{FigS1}(c,g).  Consequently, the EOM of the other variables $\{s_{x+},s_{y-},s_{z-}\}$ reduces to an effective Hamiltonian dynamics governed by the Lipkin-Meshkov-Glick (LMG) model \cite{LMG}. The long time spin dynamics in the plane of conjugate variables $\phi_i-z_i$ is regular periodic, as depicted in Fig.\ref{FigS1}(d,h). 

\textbf{Transient chaos:} In the isolated coupled-top-Dicke model ($\kappa=0$), the onset of chaos occurs above the critical point $V> V_c = J$, which can be quantified using Lyapunov exponent \cite{Strogatz}. However, when the photon loss is introduced, the vanishing of the photon field leads to synchronized dynamics between the two spins on a decoherence-free subspace, which causes the dynamics to follow that of the effective Hamiltonian of the integrable LMG model, suppressing chaos at long times. This behavior gives rise to the phenomenon of transient chaos, where chaotic dynamics persists only for intermediate times, while the asymptotic dynamics becomes regular. The transient chaos can be characterized by the early-time exponential growth and the subsequent asymptotic decay of the photon-field decorrelator (see Fig. 1(c) of the main text).

Above a critical atom-photon coupling $\lambda_c(V)$, the normal phase undergoes a transition to a superradiant phase with nonzero photon number $n\ne 0$ [see phase diagram in \ref{FigS1A}]. The stable superradiant fixed point (FSR$_1$) is given by,
\begin{subequations}
	\begin{eqnarray}
		{\rm FSR}_1: \,\,z_{1}^*\!\!&=&\!\!z_{2}^*=\!\!\sqrt{1-\frac{J^{2}(\kappa^2 +  4\omega_c^{2})^2}{\left[8 \omega_c\lambda^{2}-(\kappa^2 +  4\omega_c^{2})V\right]^{2}}},\\
		x^*\!\!&=&\!\!-\frac{8\omega_c \lambda}{(\kappa^2 +  4\omega_c^{2})} z_{1}^*, \,\, p^* = \frac{\kappa q^*}{2\omega_c},\,\, \phi_{i}^*=0. \quad
	\end{eqnarray}	
	\label{FSR1}
\end{subequations} 

%

\section{Scarring effect of unstable FSR$_2$ }
\begin{figure}[t]
	\centering
	\includegraphics[width=\linewidth]{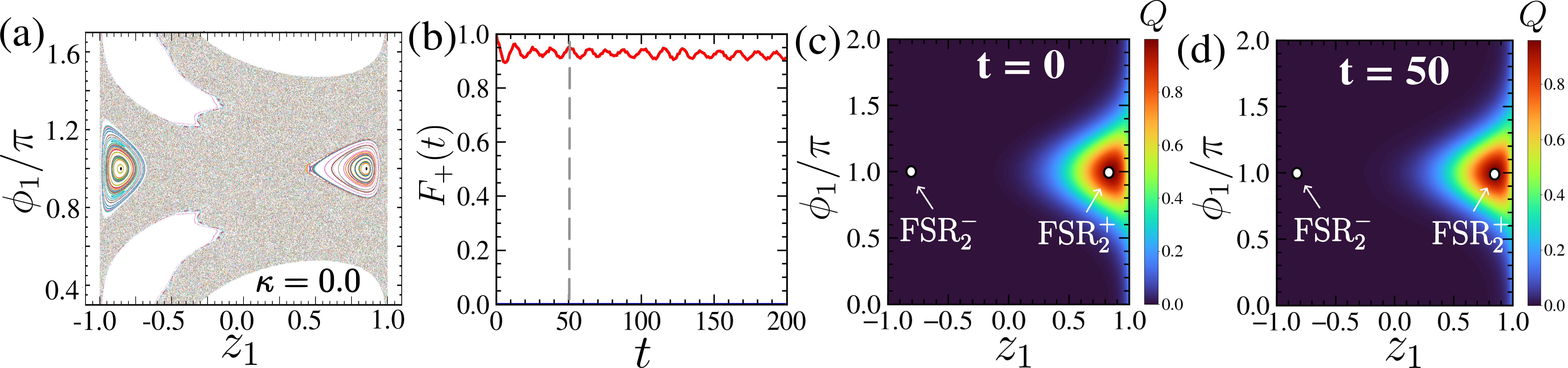}
	\caption{ Dynamical behavior around stable superradiant phase FSR$_2$ in isolated system. (a) The Poincar\'{e} section  projected on to one of the spin sectors. Regular island can be observed around the two branches of stable fixed points FSR$_2$. (b) Dynamics of the survival probability $F_+(t) = |\langle\psi(t)|\Psi_{\rm FSR_2^{+}}\rangle|^2$ starting from the coherent state $\ket{\Psi_{\rm FSR_2^{+}}}$ representing one of the branch of stable FSR$_2$. (c,d) Husimi distribution $Q(z_1,\phi_1)$ corresponding to one of the two spins in $z_1-\phi_1$ plane at two different times $t=0,50$, for initial state $\ket{\Psi_{\rm FSR_2^{+}}}$. The parameters are chosen as $\omega_c=1,\kappa=0.0,V=2.0,\lambda=0.2, S=8$.
	}
	\label{FigS6}
\end{figure}

\begin{figure}[h]
	\centering
	\includegraphics[width=0.85\linewidth]{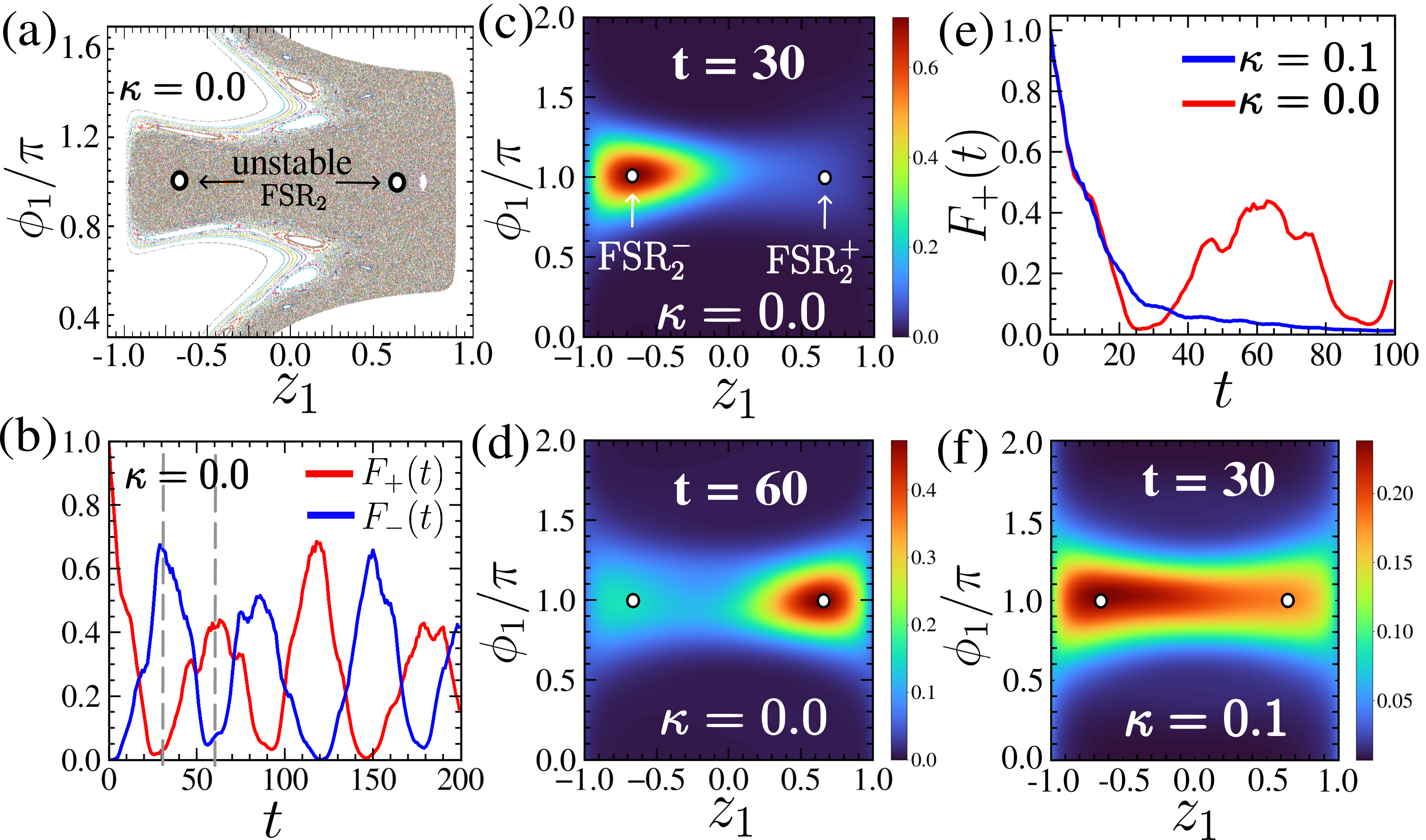}
	\caption{ Quantum scar of unstable FSR$_2$ both in isolated and dissipative systems. (a) The classical Poincar\'{e} section  projected corresponding to one of the spin sector. (b)  Dynamics of the overlaps $F_j(t) = |\langle\psi(t)|\Psi_{{\rm FSR_2}^{j}}\rangle|^2$ of time evolved state $\ket{\psi(t)}$ with coherent states $\ket{ \Psi_{ {\rm FSR_2}^{j}} }$ representing two branches of unstable FSR$_2$ with $j = \pm$ in absence of photon loss $\kappa=0$. (c,d) Husimi distribution $Q(z_1,\phi_1)$ corresponding to one of the two spins in $z_1-\phi_1$ plane at two different times $t=30,60$ for $\kappa=0$. (e) Comparison of survival probability $F_+(t) = {\rm Tr}(\hat{\rho}(t)\ket{\Psi_{\rm FSR_2^{+}}}\bra{\Psi_{\rm FSR_2^{+}}})$ between isolated (red, $\kappa=0$) and dissipative (blue, $\kappa=0.1$) cases. Note that for isolated case, $\hat{\rho}(t) = \ket{\psi(t)}\bra{\psi(t)}$. (f) Husimi distribution $Q(z_1,\phi_1)$ at time $t=30$ in presence of photon loss $\kappa=0.1$. In all panels, for quantum analysis, the initial state is the product coherent state $\ket{\Psi_{\rm FSR_2^{+}}}$, representing one of the two branches of FSR$_2$.
		Other parameters are chosen as $\omega_c=1,V=1.4,\lambda=0.2, S=8$.
	}
	\label{FigS7}
\end{figure}

To this end, we discuss the quantum scars originated as a reminiscence of the unstable superradiant phase FSR$_2$ both in the isolated system and in the presence of dissipation. As discussed in the main text, the FSR$_2$ phase exists as a stable excited state in the isolated coupled-top-Dicke model within the parameter regime indicated in the phase diagram in Fig.~1(b) of the main text. 
First, we focus on the scarring phenomenon of FSR$_2$ in the isolated system ($\kappa=0$). 
To analyze the classical phase space structure in the stable as well as in the unstable regime of FSR$_2$, we compute the Poincar\'{e} section at the fixed energy of FSR$_2$, using the surface of section defined by $p=0$. The resulting dynamics is projected on the plane of conjugate variables $z_1-\phi_1$ of one of the spin sectors.
In the stable regime, regular islands form around the two branches of fixed points, embedded within a chaotic sea, as shown in the Poincar\'{e} section in Fig.~\ref{FigS6}(a). When FSR$_2$ becomes unstable, these regular islands vanish, and the fixed points are fully submerged in the chaotic region, as depicted in Fig.~\ref{FigS7}(a).

\textbf{Signature of stable FSR$_2$ in isolated system:} To investigate the quantum mechanical signature of this superradiant phase FSR$_2$, we consider an initial product coherent state $\ket{\psi(0)} = \ket{\Psi_{\rm FSR_2^{+}}}$ representing one of the branch of FSR$_2$ and compute the survival probability 
$F_{+}(t) = |\langle\psi(t)| \Psi_{\rm FSR_2^{+}}\rangle|^2$ 
from the time evolved state $\ket{\psi(t)}$. To visualize the evolution of the phase space density corresponding to the quantum state, we also obtain the Husimi distribution $Q(z_i,\phi_i) = \frac{1}{\pi}\bra{z,\phi}\hat{\rho}_i\ket{z,\phi}$ of the reduced density matrix $\hat{\rho}_{i} = {\rm Tr}_{\bar{i}}\ket{\psi(t)}\bra{\psi(t)}$ of one of the spin sector.
In the regime of stable FSR$_2$, the survival probability $F_+(t)$ remains close to unity and decays very slowly, retaining the memory of the initial state for a sufficiently long time, as depicted in Fig.\ref{FigS6}(b). Consequently, the overlap $F_-(t) = |\langle\psi(t)|\Psi_{\rm FSR_2^{-}}\rangle|^2$ of the time evolved state with the coherent state $\ket{\Psi_{\rm FSR_2^{-}} }$ representing the other branch of FSR$_2$ remains vanishingly small.
The Husimi distribution in Fig.\ref{FigS6}(c,d) clearly displays the localized phase space density at one of the branches of FSR$_2$ with a very slow diffusion.

\textbf{Scar of unstable FSR$_2$ in isolated system:} In contrast, when the FSR$_2$ is unstable, the survival probability $F_+(t)$ decays initially and exhibits revivals [Fig.\ref{FigS7}(e)], revealing scarring phenomena. Whereas, for any generic initial states from the chaotic region, corresponding survival probability decays rapidly and vanishes.
Interestingly, a collective tunneling between the two symmetry broken branches of unstable FSR$_2$ can be observed accompanied by this scarring effect. To show this phenomenon, we compute the overlap $F_-(t) = |\langle\psi(t)|\Psi_{\rm FSR_2^{-}}\rangle|^2$ of the time evolved state with another branch of FSR$_2$, which reveals a complementary behavior from that of $F_+(t)$, as depicted in Fig.\ref{FigS7}(b). The collective oscillation of phase space density between the two branches of unstable FSR$_2$ is evident from the Husimi distribution at different times, which are shown in Fig.\ref{FigS7}(c,d).

\textbf{Scar of unstable FSR$_2$ in the presence of photon loss:} To better understand the fate of scarring phenomenon in the presence of dissipation, we compute the survival probability $F_+(t) = {\rm Tr}(\hat{\rho}(t) \ket{\Psi_{\rm FSR_2^{+}}} \bra{\Psi_{\rm FSR_2^{+}}})$ for the dissipative system ($\kappa \ne 0$) by solving master equation using stochastic wave-function method for initial state $\hat{\rho}(0) =  \ket{\Psi_{\rm FSR_2^{+}}} \bra{\Psi_{\rm FSR_2^{+}}}$ and compare it with that of isolated ($\kappa = 0$) system. Note that, in an isolated system, the states are pure, for which the $F_+(t)$ reduces to the usual definition of survival probability mentioned earlier. As evident from Fig.\ref{FigS7}(e), the initial decay of $F_+(t)$ is similar for both the dissipative ($\kappa\ne0$) and isolated cases ($\kappa = 0$); however, at longer times, the non-dissipative case exhibits revival, whereas that of the dissipative system asymptotically vanishes over the same time scale. The Husimi distribution for $\kappa \ne 0$ exhibits significantly more diffusive behavior at long times compared to the $\kappa = 0$  case, as illustrated in Fig.\ref{FigS7}(f).



\end{document}